\documentclass[3p,times,twocolumn]{elsarticle}

\usepackage{ecrc}
\usepackage{amssymb}

\usepackage{graphics}
\usepackage{graphicx}
\usepackage{epsfig}

\volume{00}

\firstpage{1}

\journalname{Physica B}

\runauth{V.E. Minakova et al.}

\jid{physb}

\jnltitlelogo{Physica B}

\usepackage{amssymb}

% ...

\begin{document}

\begin{frontmatter}


  \title{PHOTOCONDUCTION IN THE PEIERLS CONDUCTOR MONOCLINIC TaS$_3$}
  \author[label1]{V.E. Minakova\corref{cor1}}
  \author[label1]{V.F. Nasretdinova}
   \author[label1,label2]{S.V. Zaitsev-Zotov}
  \address[label1]{Kotel'nikov Institute of Radio-engineering and
Electronics of the RAS, 125009 Moscow, Russia}
  \address[label2]{Moscow Institute of Physics and Technology, 141700
Dolgoprudny, Russia }
  \cortext[cor1]{e-mail: mina\_cplire@mail.ru}





\begin{abstract}
Photoconduction in the monoclinic phase of quasi-one-dimensional conductor 
TaS$_3$ has been observed at $T < 70$~K. It was studied jointly with 
low-temperature ohmic and non-linear dark conduction. The strong sample 
quality dependence of both photoconduction and dark conduction at this 
temperature region has been
observed. Together with a similarity of the main features of the 
photoconduction characteristic of both monoclinic ({\it m-}TaS$_3$) and 
orthorhombic ({\it o-}TaS$_3$) samples the following new peculiarities of 
photoconduction in {\it m-}TaS$_3$ were 
found: 1) the dependence of the activation energy of photoconduction on 
temperature, $T$, 2) the change of the recombination mechanism from 
the linear type to the collisional one at low $T$ with a sample 
quality growth, 3) the existence of a fine 
structure of the electric-field dependence of photoconduction. 
Spectral study gives the Peierls energy gap value $2\Delta ^*= 0.18$~eV.
\end{abstract}

\end{frontmatter}

\section{Introduction}

The specific properties of the Peierls conductors
are due to charge-density-wave (CDW) formation accompanied by developing of the 
energy gap below the Peierls transition temperature $T_P$ 
\cite{MonceauREV}. 
In such materials at small electric fields the CDW is pinned 
by impurities and defects. CDW sliding starts at the 
threshold field value $E_T$ and leads to a strong non-linearity of I-V curves 
at $E > E_T$. At $E_T' < E_T$ CDW creep is possible. It causes a weak 
non-linearity of I-V curve at $E > E_T'$ which becomes clearly seen 
at low $T$ \cite{ZZcreep}. At $E < E_T'$ 
conductance is ohmic due to quasi-particles thermally exited over the Peierls 
gap.

Quasi-one-dimensional compound TaS$_3$ is known to 
have two crystal modifications, orthorhombic \cite{Bjerkelund} and monoclinic 
\cite{Rouxel1}. {\it o-}TaS$_3$ is one of the most studied Peierls conductors 
with a 
single Peierls transition at $T_P \approx 220$~K and with completely gaped 
Fermy surface. As a result, below $T_P$ the temperature dependence of the 
ohmic conductance along the chain axis, $G(T)$, obeys an activation law with 
activation energy $E_{\Delta \parallel} \approx 800$~K. It is known that $G(T)$ 
begins to deviate from the initial activation 
law at $T \lesssim T_P/2$, a new activation energy being smaller 
\cite{Takoshima}, while the 
perpendicular conductance preserves the initial value $E_{\Delta \perp} 
\approx 800$~K in all low-temperature range. 
$G(T)$ curve shape at $T \lesssim 100$~K is known to be 
highly dependent on sample 
quality -- impurity contents and degrees of structural perfection. Sometimes 
even some plateau with weakly dependent conductance bridging 
between the regions with different activation parts of $G(T)$ curve appears
\cite{Itkis,Staresinic,Ecrys12}. The origin of the longitudinal ohmic conduction 
change at low $T$ observed both in different CDW materials (blue bronze, 
(TaSe$_4$)$_2$I)  \cite{MonceauREV} and  in 
spin-density-wave compounds \cite{Gruner} is still not fully understood. The 
observation and the study of photoconduction in {\it o-}TaS$_3$ 
\cite{Pis'ma,Ecrys5,PRL} let 
us show that the main contribution into the low-temperature ohmic conduction is 
not due to quasi-particles exited over the Peierls gap, but mainly due to 
non-linear CDW excitations (such as solitons, exitons, dislocations {\it et 
al.}).

On the other hand,  {\it m-}TaS$_3$ which undergoes two Peierls transitions at 
$T_{P1} \approx 240$~K and  $T_{P2} \approx 160$~K 
is almost unstudied Peierls conductor. 
Since the first publication \cite{Rouxel1} there were only a few papers on 
this material later \cite{Itkis,Rouxel2,Roucau,Hasegawa,Maeda,ShapSt}. 
So the data were very scarce and even the main transport properties of this 
compound especially at $T < 77$~K were still 
unknown. For example, the first $G(T)$ dependence presented in \cite{Rouxel1} 
was measured for a bundle of whiskers. The second one was obtained for a single 
crystal, but the main attention was paid to the high temperature region, only 10 
points were measured at $T < 77$~K. In \cite{Rouxel2} $G(T)$ was studied only 
till 77 K. Non-linear conductance, $G(E)$, and $G(T)$ studied in 
details at $T < 77$~K in \cite{Itkis} were obtained for the only sample. $G(T)$ 
obeyed an activation law without any change of value $E_{\Delta \parallel} 
\approx 950$~K at all $T$ down to 25~K in contrast with
{\it o-}TaS$_3$ with a wide variety of $G(T)$ curve shapes at this 
temperature region. In 
\cite{Hasegawa} there was just a mention that below 50~K both $G(T)$ 
and $G(E)$ become sample dependent, $G(E)$ data were 
presented only at $T \geqslant 66$~K. So a full 
picture of the conduction at $T < 77$~K was absent. The photoconduction data  
were absent too. The main reason of a such data 
poverty is a complexity of reproducible  {\it m-}TaS$_3$ synthesis. 
The aims of this work were to fill a missing data and to compare the results 
obtained with ones available for {\it o-}TaS$_3$ 
\cite{Itkis,Ecrys12,PRL}. 

\section{Experimental}

{\it m-}TaS$_3$ crystals synthesized by the original 
technology combining the gradient technique with intermediate quenching were 
studied. The technology  
provides reproducible synthesis of {\it m-}TaS$_3$ with $\approx$ 100~\% content 
of monoclinic phase crystals in the batch. Eight investigated samples from two 
batches were not so extremely thin as 
{\it o-}TaS$_3$ ones studied in \cite{Ecrys12,Pis'ma,Ecrys5,PRL}. The typical 
transverse sample sizes were (1-10)~$\mu$m. 
The measured heating effect was $\lesssim 3$~mK due to excellent thermal 
contact with sapphire substrate.

Since both the dark conduction and photoconduction turned out to be highly 
dependent on a sample quality, we tried to use in our investigations 
{\it m-}TaS$_3$ samples of different quality in order to get a 
complete picture of the phenomenon. The quality was estimated in two ways: 1) by
the measurement of electric-field dependent dark non-linear conductance at 
$10$~K $< T < 140$~K to determine $E_T$ and $E_T'$ values; 
2) by the examination of dark ohmic conductance, $G(T)$, measured in a wide 
temperature range $10$~K $ < T < 300$~K at sample voltage $V \lesssim V_T'$.

Three types of photoconduction study were undertaken. 1) Investigation of the 
temperature dependences of photoconductance, $\delta G(T) = G(T,W) - G(T,0)$, 
at various light intensity levels, $W$, at $V \lesssim V_T'$. 
2) Combined  study of the electric-field dependences of non-linear dark 
conductance, $G(E)$, and 
of photoconductance, $\delta G(E)$, at different $T$ and $W$. 3) 
Photoconduction spectroscopy for two types of the incident light 
polarization (along and transverse to the crystal chain axis) at 
photon energy range (0.1 - 0.45)~eV at different $T$. 

All conductance measurements were done along the chain direction in two-probe 
configuration in the voltage-controlled regime. A resistance of contacts 
made by indium cold soldering was $\sim$ 10 $\Omega$. IR LED 
(light flux $W = (10^{-4}$ - 
$30)$~mW/cm$^2$ at the sample position, a wavelength $\lambda = 0.94$~$\mu$m) 
driven by meander-modulated current was used. The photon energy 
1.3 eV was $> 2\Delta ^*$. (For {\it o-}TaS$_3$ $2\Delta ^*= 0.25$~eV at $T = 
40$~K \cite{zzhere}.) Photoconduction signal was measured 
by the usual AC modulation method at modulation frequency $f = 4.5$~Hz. 
Photoconduction spectra were measured with a use of a grating 
monochromator (evacuated to a pressure below 1 Torr to reduce the effect of 
light absorption by the air) with a globar, $f = $3.125~Hz. 
Other measurement details are in 
\cite{Ecrys12,Pis'ma,Ecrys5,PRL,Nasr}.

\begin{figure}
\includegraphics[width=7.5cm]{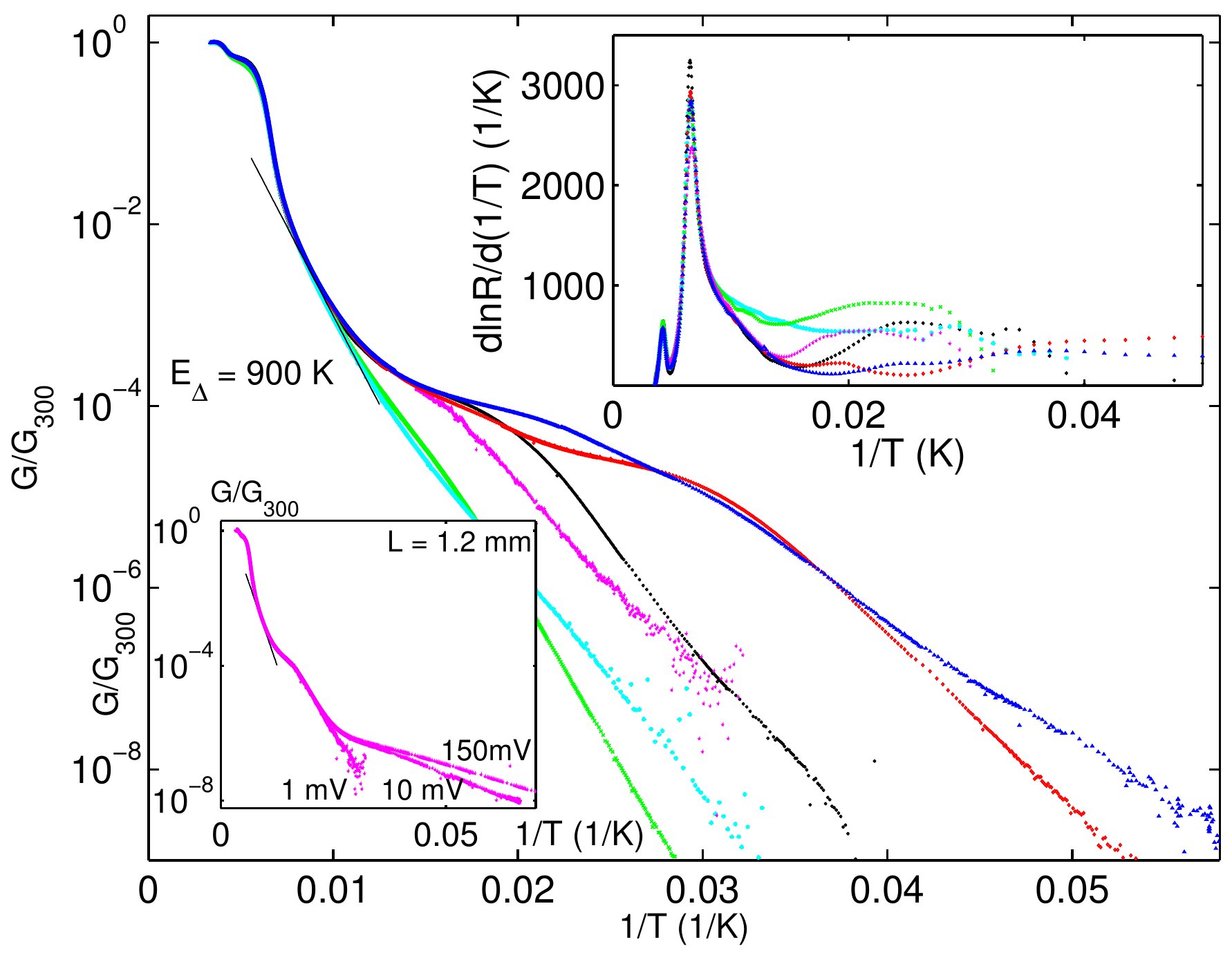}
\caption{(Color online)Temperature dependences of dark ohmic conductance, 
$G(T)$, normalized 
to its room-temperature values, $G_{300}$, for different {\it m-}TaS$_3$ 
samples. 
Top inset shows corresponding temperature derivatives of the resistance, 
$d\ln R/d(1/T)$. Bottom inset shows $G(T)$ curves measured at various 
sample voltages for the sample marked by magenta color in the main figure.}
\label{Fig.1}
\end{figure}

\begin{figure}
\includegraphics[width=7.5cm]{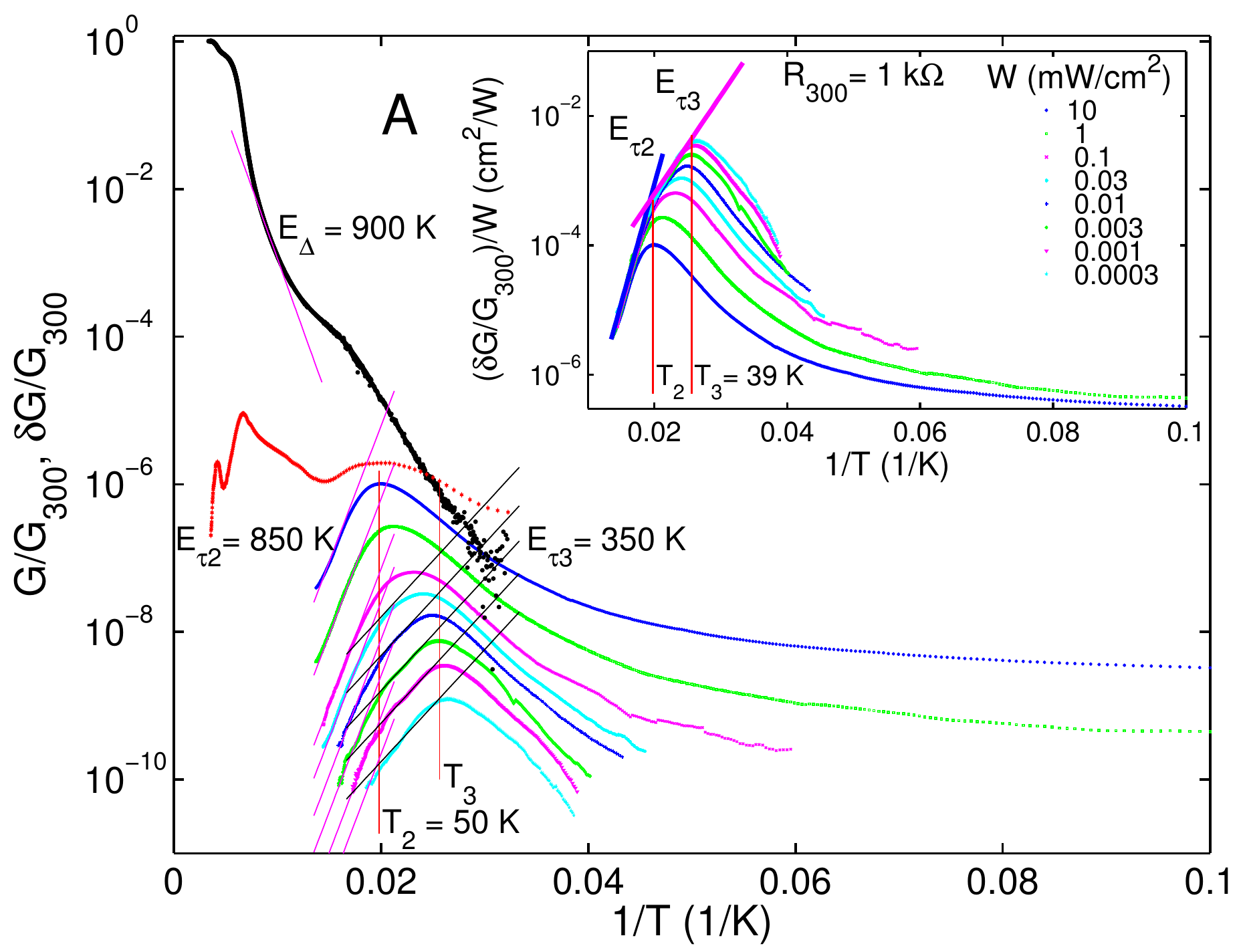}\\
\includegraphics[width=7.5cm]{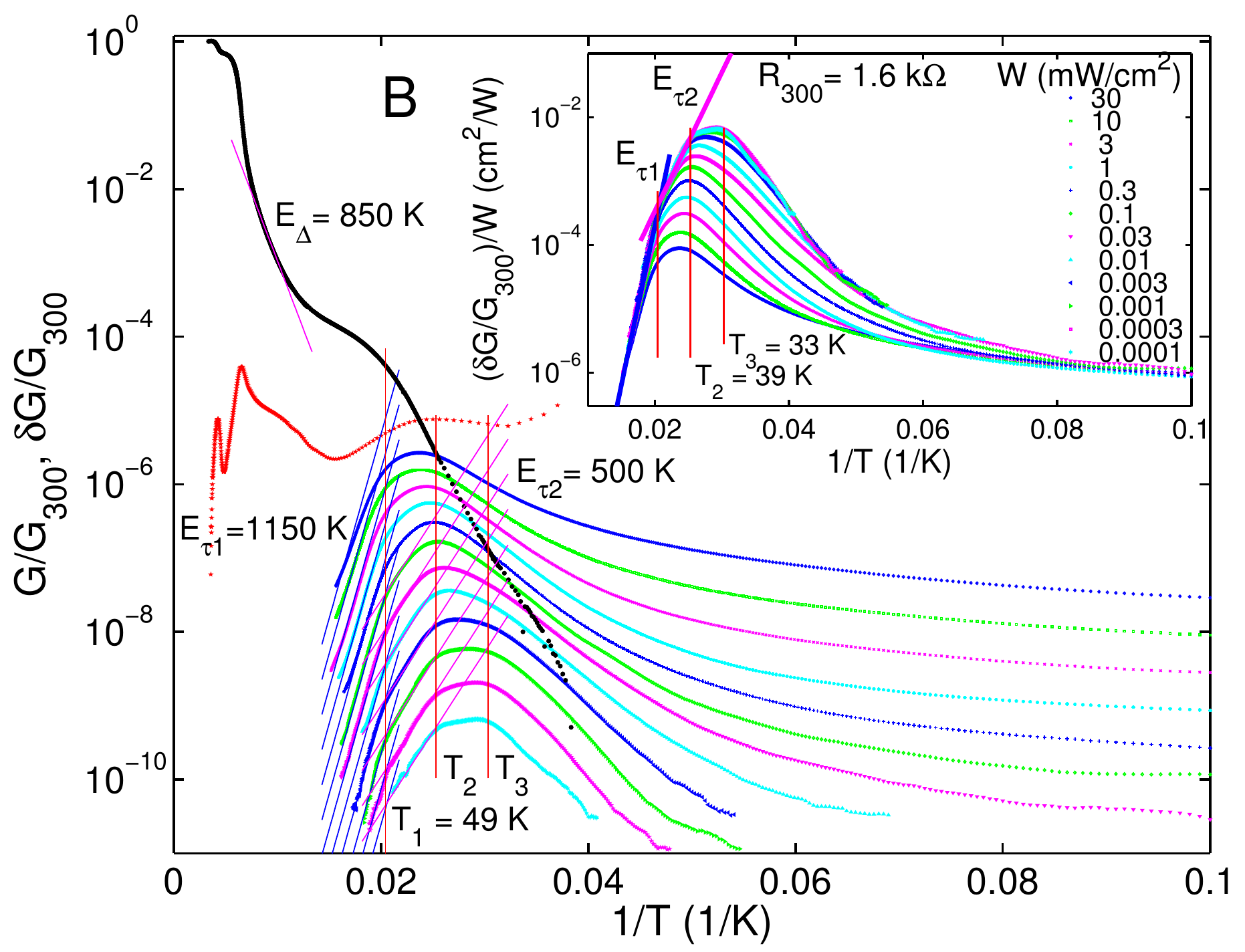}\\
\includegraphics[width=7.5cm]{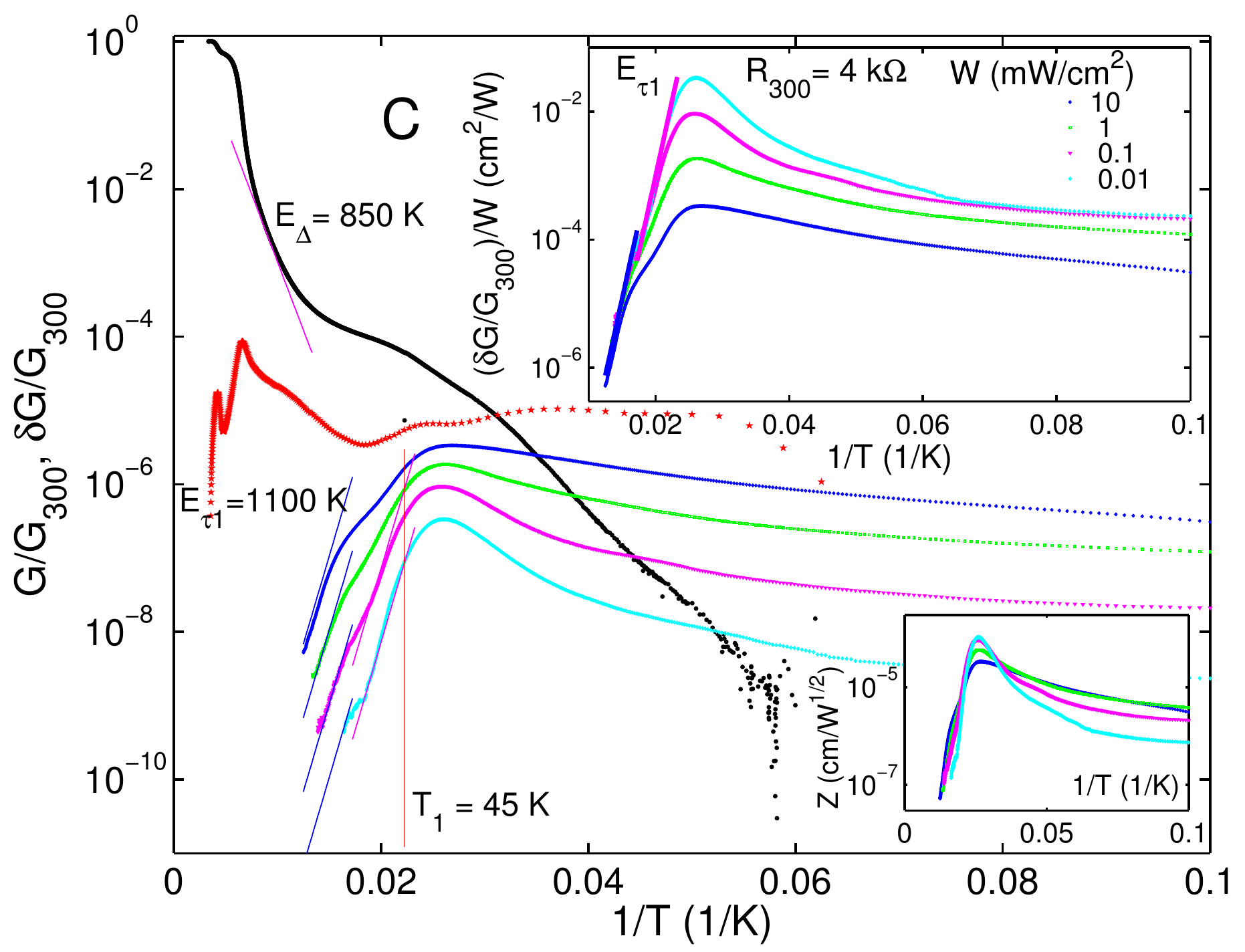}
\caption{Sets of normalized temperature dependences of photoconductance, 
$\delta G(T)/G_{300}$, at different light intensities, $W$, together with 
normalized dependences of 
dark ohmic conductance, $G(T)/G_{300}$, (black curves) and 
temperature derivatives of the resistance, $d\ln R/d(1/T)$, (red curves, 
arb. units) 
for the samples of the groups A, B and C, respectively (see text). The 
top insets show the same dependences $\delta G(T)/G_{300}$ normalized to 
corresponding values of $W$. The bottom inset shows the same dependences 
normalized to corresponding $\sqrt W$ values: $Z(T) = (\delta 
G(T)/G_{300})/\sqrt W$.}
\label{Fig.2}
\end{figure}
\section{Temperature dependences of the ohmic conductance and of the
photoconductance}

 Fig.~\ref{Fig.1} shows the dark $G(T)$ dependences normalized to its 
room-temperature values, $G_{300}$, together with corresponding temperature 
derivatives of the resistance, $d\ln R/d(1/T)$, (in the top inset) of {\it 
m-}TaS$_3$ samples. 
At $T \gtrsim 77$~K the difference between the curves is 
small. For all the curves the well-defined activation region just below 
$T_{P2}$ is absent, the activation energy $E_{\Delta \parallel} = 850 - 900$~K 
is 
temperature dependent, 
$T_{P1} \approx 238$~K, $T_{P2} \approx 153$~K. The situation changes at $T 
\lesssim 77$~K: a huge difference of $G(T)$ shapes analogous to one 
for {\it o-}TaS$_3$ appears. For low-quality samples a plateau in 
$\log G(1/T)$ curve at $T < 77$~K as a rule is small or is absent at all. With 
a sample quality growth the plateau width generally increases.
A wide maximum is clearly seen in $d\ln R/d(1/T)$ curve in this temperature 
region.

Six samples with different plateau width were chosen for the further 
investigations. To provide correct measurements of $\delta G(T,W)$ amplitude 
both at the high and at the low temperature regions (where the signal is very 
small) the dependences 
$\delta G(T)$ at all $W$ were measured at both signs of the voltage 
bias for excluding LED current modulation crosstalk.
The samples which were measured at the same conditions were divided into three 
groups (A, B and C, each of two samples) according with the shape of $G(T)$ 
curve, see Fig.~\ref{Fig.2}: small plateau width - A, large plateau width - B, 
huge plateau width - C. All figures below correspond to the same samples from 
the each group.

The common photoconduction features of {\it m-}TaS$_3$ and {\it o-}TaS$_3$ 
\cite{Ecrys12,PRL} are the following: 

1) Photoconduction is detected only at $T \lesssim T_P/2$.

2) The $\delta G(T)$ dependences are non monotonic with a maximum, its 
magnitude and position, $T_{max}$, being dependent on $W$. 

3) For the samples with a huge plateau region two maxima of $\delta G(T)$ may 
take place (group C) \cite{Ecrys12}.

4) There is a correlation between $G(T)$ and $\delta G(T)$: for the
samples with a wider plateau the $G(T)$ value at a fixed $T$ is larger, the 
$\delta G/G_{300}$ signal at the maximum at fixed $V$ and $W$ values is larger 
too (group C, B). Moreover, both $\delta G(T,W)$ maxima and a wide 
low-temperature $d\ln R/d(1/T)$ maximum 
(red curve in Fig.~\ref{Fig.2}) are approximately at the same $T$. $T_{max}$ 
decreases with the plateau width growth, $E_T$ value as a rule reduces too.

5) $\delta G(W)$ dependence is linear at $T > T_{max}$. As a result, $\delta 
G(T)$ curves at $T > T_{max}$ obey the unique activation law for all $W$, 
the activation energy ($E_{\tau 1}$ for groups B, C) being 
approximately the same for good-quality {\it m-}TaS$_3$ and {\it o-}TaS$_3$ 
samples. ($E_{\tau}$ value is determined by the current carrier recombination 
time which obeys an activation law in this temperature region 
\cite{Ecrys12,PRL}.)

6) The activation energies in both materials are smaller for the 
low-quality samples ($E_{\tau 2}$ for group B).

New photoconduction and dark ohmic conduction features of 
{\it m-}TaS$_3$ undetected in {\it o-}TaS$_3$ are observed: 

1) There are several different temperature regions in the high temperature part 
of $\delta G(T)$ curves (two regions for the group A samples and three 
regions for the group B ones) with different activation energies of $\delta 
G(T)$, the 
low-temperature activation energy being 
smaller. Unfortunately we have no data for the representative group C sample at 
sufficiently small $W$ levels.

2) The region with the highest activation energy, $E_{\tau 1}$, is absent for 
the group A samples.

3) In these temperature regions a linear recombination regime takes place for 
all sample groups. As 
a result, $\log \delta G(1/T)$ curves normalized to the corresponding $W$ values
coincide in the regions with different slopes, changing with $T$ decrease 
(top insets in Fig.~\ref{Fig.2}).

4) For the low-temperature region the recombination type is different for 
the different sample groups. For the low-quality samples $\log \delta G(1/T)/W$ 
curves coincide also at $T \lesssim 10$~K, indicating the linear type of 
recombination. The collisional recombination regime is not achieved 
for the group A samples at all available $W$. 
With a sample quality growth a region with a difference between these curves 
(near and just below $T_{max}$) narrows, $(\delta G/G_{300})/W$ value 
at low $T$ becomes higher, and at last the difference between the curves at low 
$T$ appears. For the group C (and for some of the group B) samples the linear 
recombination regime takes place at low $T$ only for a small $W$. With $W$ 
growth the situation changes: the coincidence of $\delta G(T)$ curves normalized 
to corresponding $\sqrt W$ values for large $W$ at $T \lesssim 30$~K (bottom 
inset in Fig.~\ref{Fig.2}, C) indicates a transition into
collisional recombination regime. So, we can conclude: the 
higher the sample quality is the 
smaller $W$ values are required for the change of the recombination mechanism 
and for the realization of the collisional recombination regime at low $T$. 
This regime was also observed in pure {\it o-}TaS$_3$ samples at $T \lesssim 
30$~K \cite{PRL}. 

5) For the samples with two maxima in $\delta G(T)$ (group C) the 
plateau region in $G(T)$ may split on two parts (Fig.~\ref{Fig.1}, red curve), 
and the positions of $d\ln R/d(1/T)$ curve peaks correlate with ones of 
$\delta G(T)$ maxima.

6) The height of the low-temperature maximum in $\delta G(T)$ may be both less and 
larger than the height of the high-temperature one.

\section{Electric-field dependences of the photoconductance and of the dark 
conductance}
\begin{figure}
\includegraphics[width=7.5cm]{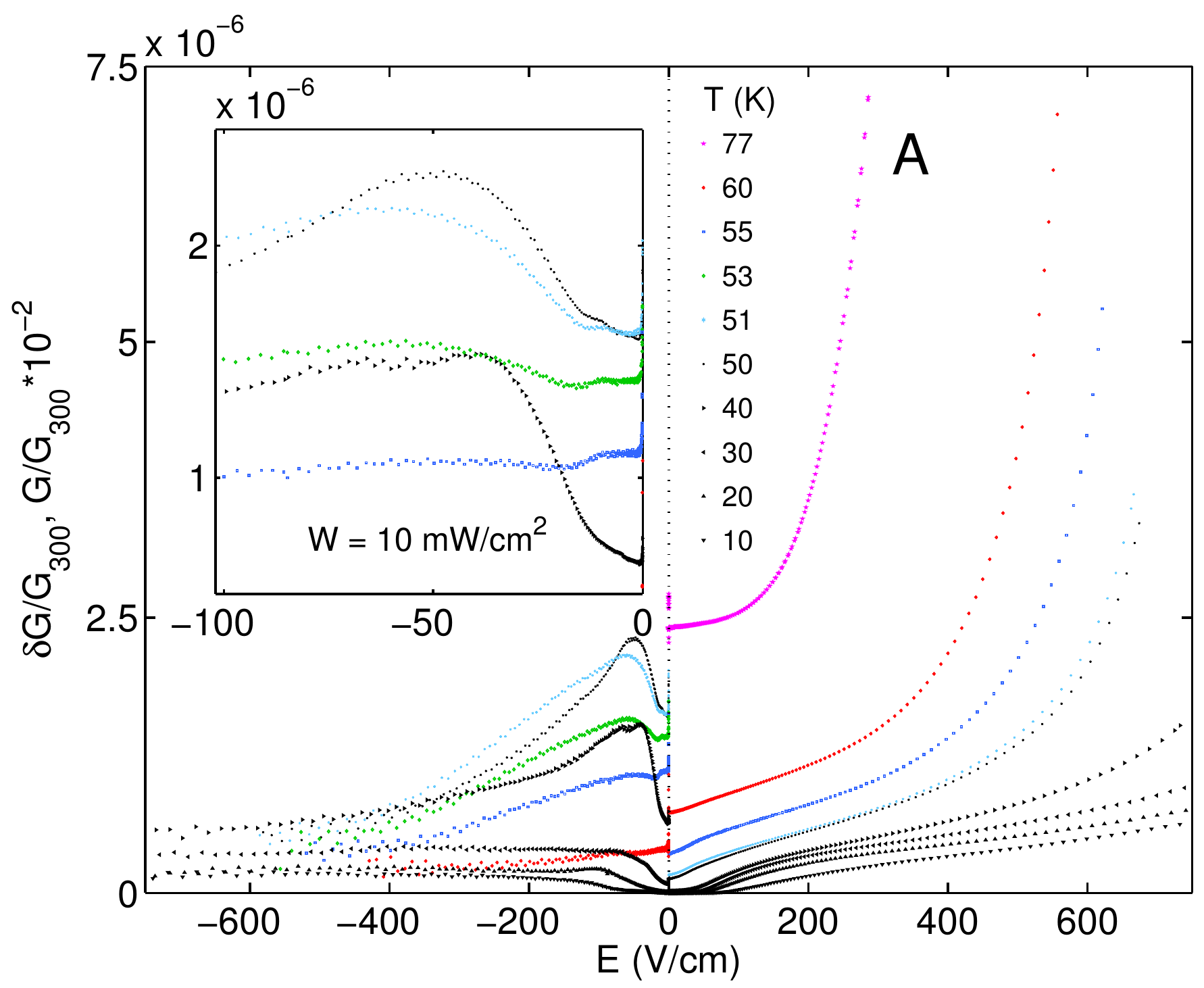}\\
\includegraphics[width=7.5cm]{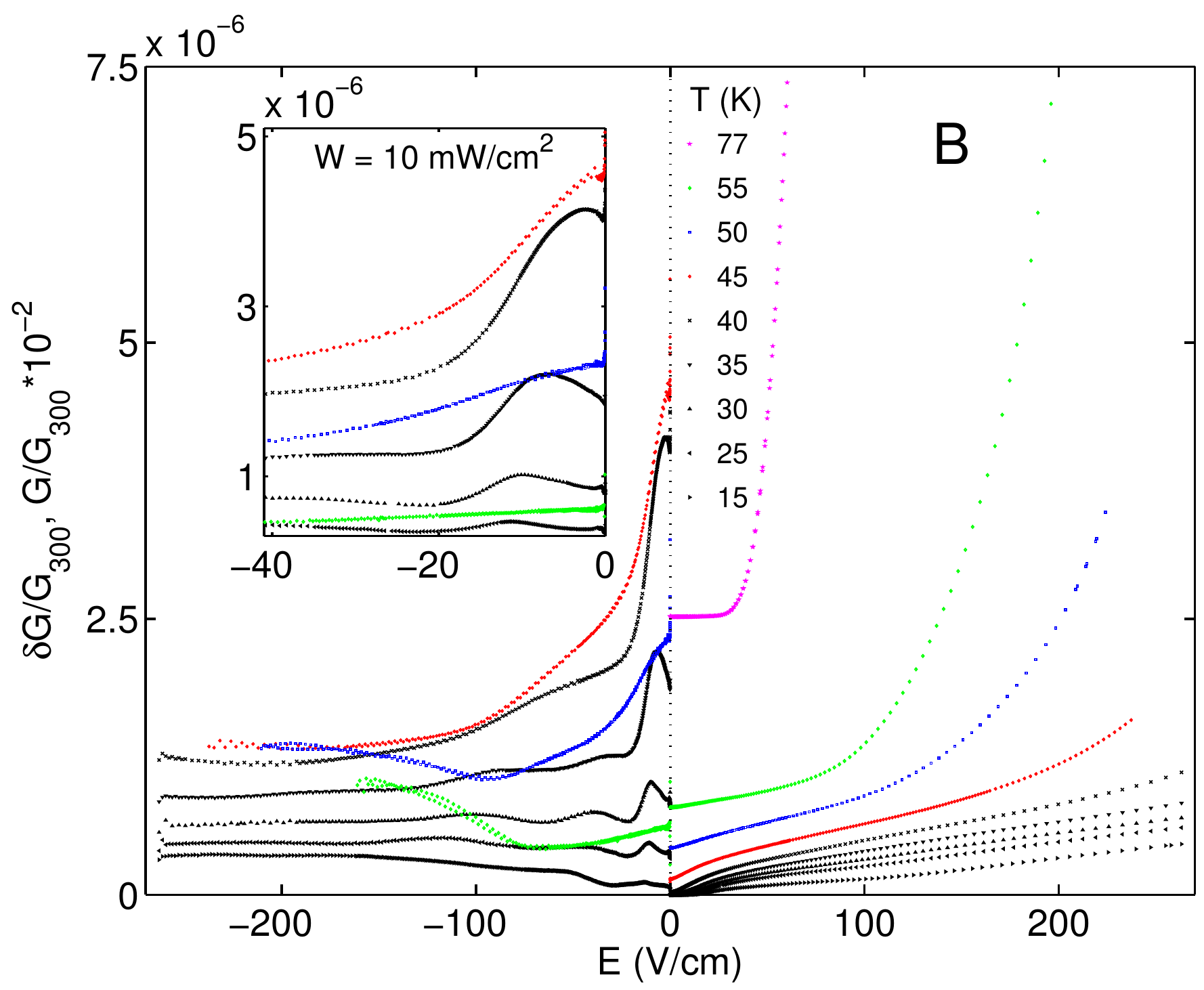}\\
\includegraphics[width=7.5cm]{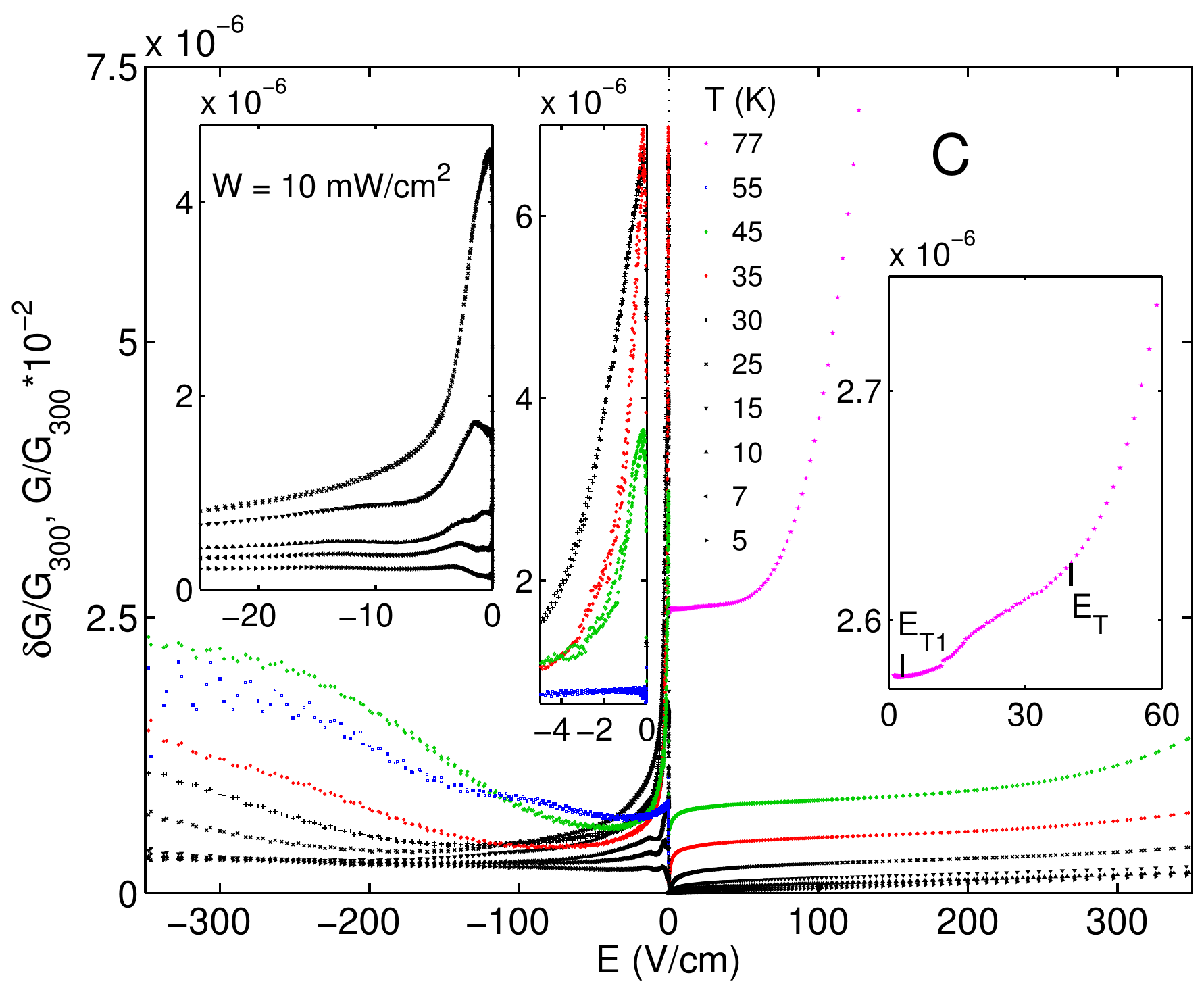}
\caption{Sets of normalized electric-field dependences of the dark conductance, 
$G(E)/G_{300}$, (in the right) and of the corresponding normalized 
photoconductance, $\delta G(E)/G_{300}$, at a fixed $W$ (in the left) at 
different $T$ for all the sample groups. 
Color curves are measured at $T > T_{max}$, black ones -- at $T < T_{max}$. 
Left and right insets show the initial parts of $\delta G(E)$ and $G(E)$ 
curves, respectively.}
\label{Fig.3}
\end{figure}
\begin{figure}
\includegraphics[width=7.5cm]{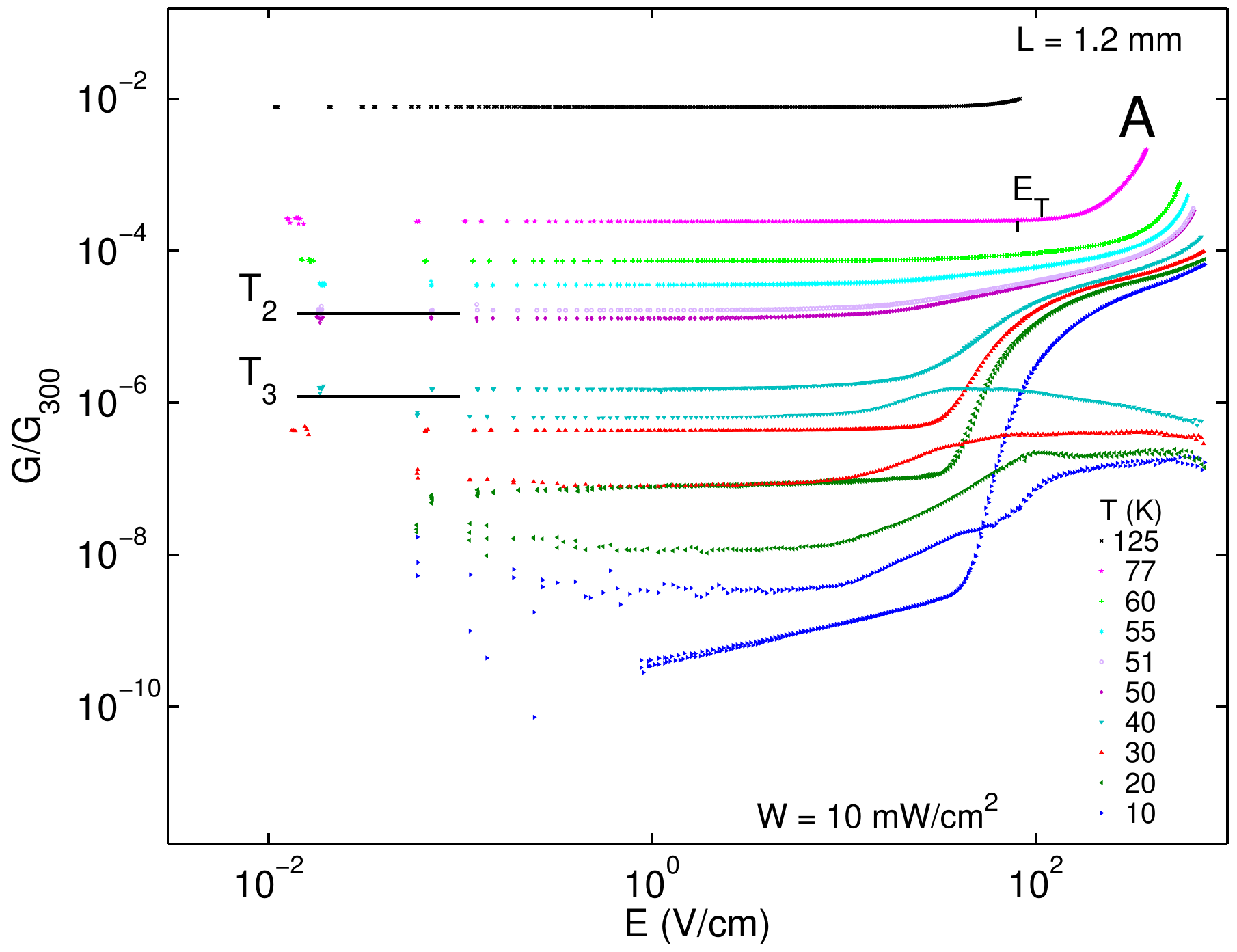}\\
\includegraphics[width=7.5cm]{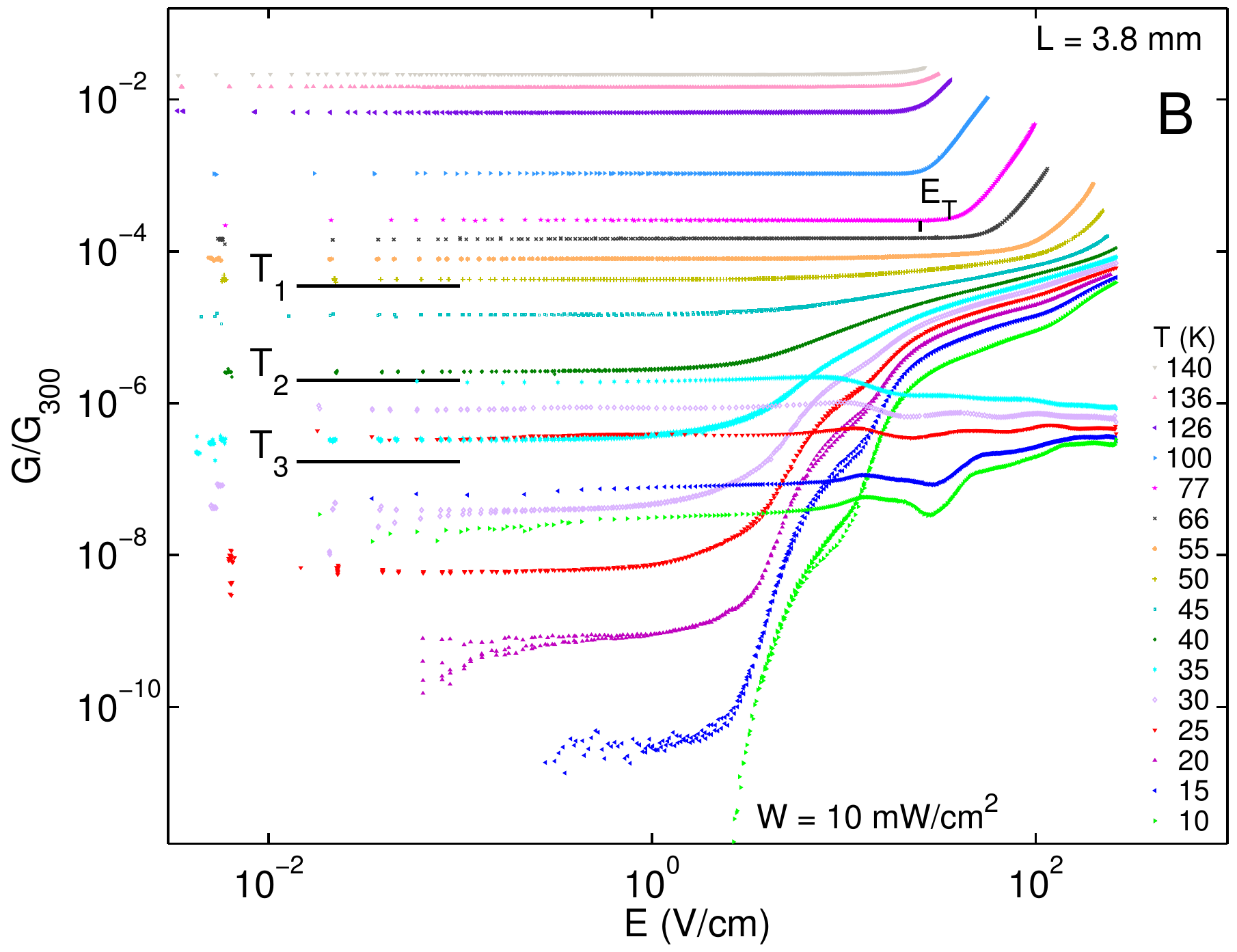}\\
\includegraphics[width=7.5cm]{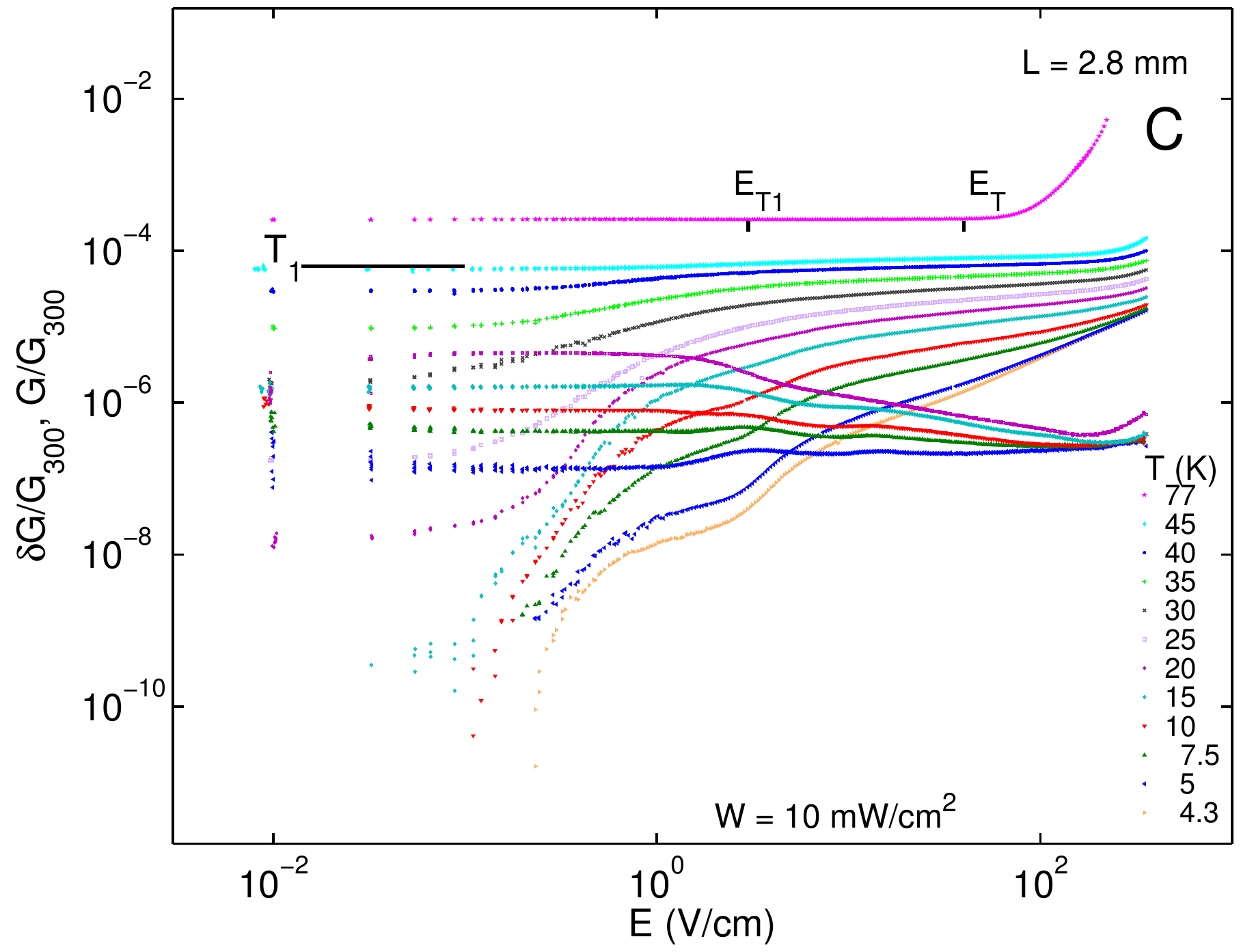}
\caption{Normalized electric-field dependences of the dark conductance,  
$G(E)/G_{300}$, (upper 
set of data) and of the photoconductance, $\delta G(E)/G_{300}$, (lower 
set of data) at different $T$ for all the sample groups.} 
\label{Fig.4}
\end{figure}
Fig.~\ref{Fig.3} shows sets of normalized $\delta G(E)$ curves collected 
at a fixed $W$ (in the left) 
along with corresponding dark $G(E)$ curves (in the right) at 
different $T$ for the all groups. $G(E)$ measurements were made before the 
illumination to exclude residual photoconductance after turn-off of the light.
All the curves were measured after application of voltage $V > V_{T}$ to remove a 
metastability causing a strong hysteresis of the curves. 
Sample quality growth reduces the hysteresis.
In Fig.~\ref{Fig.3} the color curves were measured at $T > T_{max}$, black ones 
-- at $T 
< T_{max}$. 
Although the minimum of $E_T$ value for our samples is achieved at $T \approx 
130$~K, $G(E)$ at $T = 77$~K (magenta curves in Fig.~\ref{Fig.3},~\ref{Fig.4}) 
are presented as a sample specification, the photoconduction is yet not seen.

The common features of photoconduction and dark conduction for all the sample 
groups are the following: 

1) Both $\delta G(E)$ and $G(E)$ dependencies are highly non-linear. 

2) The dark conductance decreases monotonously with a decrease of $T$. 

3) The photoconductance change is non-monotonic with a decrease of $T$. Both 
an evolution and a shape of the color $\delta G(E)$ curves in Fig.~\ref{Fig.3} 
differ from ones of the black curves. With a decrease of $T$ the color 
$\delta G(E)$ curves rise up and the peak at $E = 0$ increases. 
With a further $T$ decrease black $\delta G(E)$ curves wholly come down, the 
peak splits and becomes smaller. 

4) In whole, a $\delta G(E)$ curve shape for all the groups is the similar. 
Almost for all $T$ photoconductance decreases with an increase of $E$ 
then for sufficiently large $E$ values (group C, low $T$ or all 
groups, high $T$) it begins to growth. At $T 
\lesssim 15$~K a growth of $\delta G(E)$ is remarked.

The main differences of $\delta G(E)$ and of $G(E)$ dependences  
for various quality samples are the following:

1) The peak at $E = 0$ dominates in $\delta G(E)$ color curves for the samples 
of the groups B and C, whereas for the group A the peak at $E = 0$ is of 
the same value or even less than neighbor one (left insets in Fig.~\ref{Fig.3}).

2) The peak is higher and more narrow for more quality samples. 

3) A fine structure of $\delta G(E)$ dependences consisting of a few 
additional peaks develops with $T$ decreasing. It is clearly seen at low 
$T$ for groups B and C.

4) For group A the dependences $G(E)$ are nonlinear at $T 
\leqslant 77$~K for all $E$. Even at $T = 77$~K a weak nonlinearity 
is preceded by the strong one. The abrupt threshold for the onset of CDW sliding 
is absent. A sufficiently sharp threshold for the group B sample 
and two-threshold behavior of $G(E)$ curve (with $E_{T1}$ and $E_{T}$ values) 
for group C sample occur at $T = 77$~K.

The sample difference is more clearly seen in Fig.~\ref{Fig.4}, which shows the 
same sets of $G(E)$ curves in a double logarithmic scale. $E_T$ value of 
good-quality samples is a few times smaller than that of the group A. 
$E_T$ increases with a decrease of $T$ for all the groups. 
It is difficult to determine distinctly the onset of a weak nonlinearity, 
$E_T'$, 
at low $T$ for all the samples. We can only say that $E_T'$ value reduces with 
a decrease of $T$ and becomes too small at low $T$. 
As a result, at $E > E_T'$ an additional conduction channel turns on at $T 
\lesssim 50$~K. 
Account of this fact is very important at $G(T)$ measurements at low $T$. 
A small increase of a sample voltage leads to dramatic changes of $G(T)$ 
shape: an additional $G(T)$ curve deviation appears at low 
$T$, see bottom inset in Fig.~\ref{Fig.1} (group A sample). In this regime the 
low-temperature $G(T)$ tail obeys an activation law, the activation energy 
being dependent on sample voltage. Such magenta curves 
measured at large voltages have a shape similar to ones for 
{\it o-}TaS$_3$ samples in \cite{Itkis}. The fact that the same $G(T)$ curve at 
$V = 1$~mV does not changes and has an activation behavior at least till $T = 
35$~K let us estimate the upper boundary of $E_T'$ value for the group A sample 
at $T = 35$~K. It is equal to 0.008~V/cm. Analogously, $E_T' \leqslant 
0.05$~V/cm at $T = 30$~K for the group B, $E_T' \leqslant 0.05$~V/cm at 
$T = 20$~K for the group C. 
As it is known from the photoconduction study of {\it o-}TaS$_3$ \cite{Pis'ma} 
the appearance of $G(E)$ nonlinearity reduces the current carrier recombination 
time due to speed-up of recombination of spatially separated electrons and holes 
and leads to a decrease of photoconduction response. 

Fig.~\ref{Fig.4} also helps to reveal invisible in Fig.~\ref{Fig.3} appearance 
and evolution of a fine structure of $G(E)$ dependences with $T$ decreasing, 
which is clearly seen for the groups B and C and is invisible for the group A.

Fig.~\ref{Fig.4} also shows $\delta G(E)$ curves at low $T$. One can see some 
correlations between fine 
structures of $\delta G(E)$ and $G(E)$ curves: 1) the both types of 
fine structures exist in the same $E$ and $T$ regions, 2) the more explicitly 
the $G(E)$ fine structure is seen the more 
visible the fine structure in $\delta G(E)$ curves becomes.
It seems the nature of the both fine structures is the same. Apparently,
$\delta G(E)$ fine structure may be also seen in {\it o-}TaS$_3$.

\section{Photoconduction spectroscopy}
\begin{figure}
\includegraphics[width=7.3 cm]{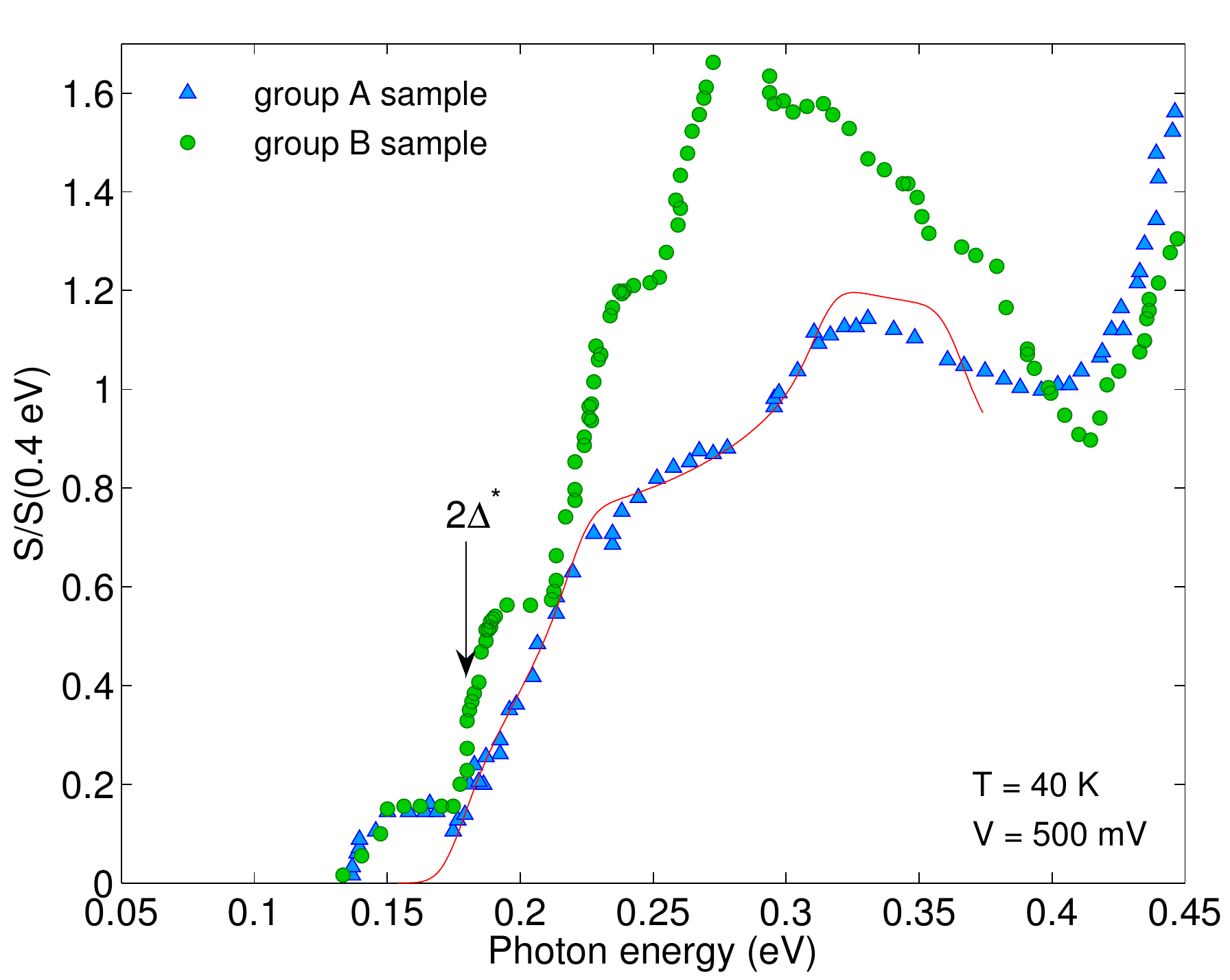}
\caption{Photoconduction spectra of the samples of the groups A and B.
Solid curve shows the best fitting by the Peierls gap modulation model 
\protect\cite{zzhere}.}
\label{Fig.5}
\end{figure}

Currently, photoconduction spectra were 
measured only for the samples of  the groups A and B. 
The data obtained at $T = 40$~K and voltage $V = 500$~mV for both 
samples are shown in the Fig.~\ref{Fig.5}, where $S(\hbar \omega)=\delta 
G/\hbar \omega$. The spectra shape reminds that of {\it o-}TaS$_3$ 
\cite{zzhere,Nasr} except for a wide peak at 0.15 eV. The data look 
more reproducible than those in {\it o-}TaS$_3$ except for an  
additional step-like structure  (periodical oscillations with $\approx 50$~meV 
periodicity) seen on the group B sample data. The measured Peierls energy gap 
value, $2\Delta ^*= 0.18$~eV.

\section{Concluding remarks}
The studied {\it m}-TaS$_3$ samples have been  
divided into three groups depending on plateau width in $G(T)$ curve.
The samples sizes were approximately the same. The experimental conditions 
were also the same for all the samples. All the measured dependences for these 
groups turned out to be different.
So, the reason of the difference between all the sample groups is mainly 
the sample quality, which is determined by two factors: 
1)~impurity contents, 2) degrees of structural perfection.

The impurity contents were different:
the low-quality samples ($E_T > 25$ V/cm) -- group A and
good-quality samples ($E_T$ = 3 - 15 V/cm) -- groups B and C.
Unfortunately we had no high-quality samples ($E_T \lesssim 1$~V/cm).
So the group A samples differ from the group B and C ones 
firstly due to impurity contents. The presence of impurities kills all fine 
effects -- the fine structure of $G(E)$ and $\delta G(E)$, step-like structure 
of the spectrum, high temperature part of $\delta G(T)$ curve characterizing by 
the highest activation energy value, $E_{\tau 1}$. Thus the presence of 
impurities reduces the current carrier recombination time due to providing an 
additional channel of non-equilibrium current carrier recombination and leads 
to an impossibility of the realization of the collisional recombination regime 
at low $T$.

As a rule, the plateau width is smaller for the low-quality samples both for 
{\it o}-TaS$_3$ and {\it m}-TaS$_3$ case.
But the presence of impurities is not the only reason of the plateau reduction. 
For example, we have met {\it o}-TaS$_3$ samples doped by Nb (concentration 
$\sim 0.5$\%) with a huge plateau. 
In our opinion the additional reason of the difference between the groups is 
due to different 
degrees of structural perfection, which we unfortunately can not control. 
It seems a big plateau is caused by a structural 
inhomogeneity of the crystal. Two different types of the inhomogeneity (which 
we can not distinguish yet) may take place. 
1) A sample may be an aggregate of a few parallel crystals with different $E_T$ 
values (the longitudinal stripe structure is often seen in thick 
{\it o}-TaS$_3$ crystals). 2) A sample may be deformed by outside factors -- for 
example by stretch 
arising in the process of contacts fabrication. Our preliminary results 
obtained for {\it o}-TaS$_3$ crystals with different degrees of the 
longitudinal stretch confirm this assumption. It seems impurities 
prevent a sample from outside deformation, it leads to the reduction of the 
stretch possibility of the low-quality~samples. 

It also worth to mention that $G(E)$ dependencies of our samples in general are 
consistent with the same ones for high 
quality {\it m-}TaS$_3$ sample reported in \cite{Itkis}. Moreover the same fine 
structure is also present but was not remarked and discussed in \cite{Itkis}. 
The similar peculiarities of IV-curves were also observed and analyzed in impure 
samples of {\it o-}TaS$_3$ at $T \lesssim 10$~K \cite{Waves} where they were 
attributed to the difference of the pinning potentials provided by various 
impurities. 
Here such a difference may also come from separate depinning of two CDWs. 

Another interesting observation is very strong and unusual $\delta G(E)$ 
dependence. 
The concurrence  of two physical mechanisms may result in such a complex 
behavior. 
On the one hand, spatial separation of electrons and holes caused by 
fluctuations of the chemical potential (which are significant at low $T$ 
\cite{Pok}) 
complicates their recombination process \cite{Pis'ma,Ecrys5}. However the 
process accelerates with CDW motion. 
On the other hand, with a decrease of $T$ a role of electric-field 
dependent electronic states with relatively small excitation energies
(impurities, defects, excitons {\it et al.}) which can promote the 
non-equilibrium current carrier recombination increases.
Our photoconduction spectroscopy data confirm the existence of the 
electric-field dependent electronic 
states in {\it o}-TaS$_3$ \cite{Nasr}.

The group A spectrum shape can be reasonably fitted (solid curve in 
Fig.~\ref{Fig.5}) within 
a simple model of Peierls gap modulation \cite{zzhere} except for a small and 
wide peak around 0.15 eV. This spectrum exhibits van Hove 
singularities arising due to the manifestation of the 3D properties of 
the compound. The detail discussion of the origin of van Hove singularities 
in both o-TaS$_3$ and m-TaS$_3$ case one can find in \cite{zzhere}. 
Surprisingly, no signature of the second Peierls gap 
is seen at $\hbar \omega<0.4 eV$. The group 
B spectrum shape is very similar but 
complicated by an additional periodical step-like structure. The period of 
these oscillations ($\approx 50$~meV) is too big to be attributed to the light 
interference, so it is more likely corresponds to 
the phonon repetitions of the peak 0.15 eV. Similar oscillatory structure has 
been also found in {\it o-}TaS$_3$.

The main results of photoconduction and of low-temperature conduction 
study in Peierls conductor {\it m-}TaS$_3$ are similar 
to ones in {\it o-}TaS$_3$. However much more details are observed in {\it 
m-}TaS$_3$. The differences are believed to be a consequence of more perfect 
crystal structure of {\it m-}TaS$_3$ 
in comparison with {\it o-}TaS$_3$. For example, it is known that the RF-field 
response in {\it m-}TaS$_3$ also is more pronounced than that in {\it 
o-}TaS$_3$ due to more higher CDW coherence in monoclinic phase of the compound 
\cite{ShapSt}. So, 
{\it m-}TaS$_3$ is very grace object of research.
The obtained phenomenon picture is very complex and needs further 
investigations, which as we hope let us clarify the nature of 
low-temperature conduction in CDW materials.\\

{\bf Acknowledgements.} The work was supported by RFBR project 14-02-01236 and 
Department of physical science of RAS.



\end{document}